\documentclass[%
 reprint,
 superscriptaddress,
 showpacs,preprintnumbers,
 amsmath,amssymb,
 aps,
 prl,
floatfix
]{revtex4-1}

\usepackage{graphicx}% Include figure files
\usepackage{dcolumn}% Align table columns on decimal point
\usepackage{bm}% bold math
\usepackage[colorlinks,allcolors=blue]{hyperref}% add hypertext capabilities
\usepackage{txfonts}
\usepackage{pdfpages}

\begin{document}

\preprint{}

\title{Temperature-dependent phonon spectra of magnetic random solid solutions}

\author{Yuji Ikeda}
\email{ikeda.yuji.6m@kyoto-u.ac.jp}
\affiliation{Center for Elements Strategy Initiative for Structure Materials (ESISM), Kyoto University, Kyoto 606-8501, Japan}

\author{Fritz K\"{o}rmann}
\affiliation{Department of Materials Science and Engineering, Delft University of Technology, Mekelweg 2, 2628 CD Delft, Netherlands}
\affiliation{Max-Planck-Institut f\"{u}r Eisenforschung GmbH, D-40237 D\"{u}sseldorf, Germany}

\author{Biswanath Dutta}
\affiliation{Max-Planck-Institut f\"{u}r Eisenforschung GmbH, D-40237 D\"{u}sseldorf, Germany}

\author{Abel Carreras}
\affiliation{Department of Materials Science and Engineering, Kyoto University, Kyoto 606-8501, Japan}

\author{Atsuto Seko}
\affiliation{Center for Elements Strategy Initiative for Structure Materials (ESISM), Kyoto University, Kyoto 606-8501, Japan}
\affiliation{Department of Materials Science and Engineering, Kyoto University, Kyoto 606-8501, Japan}
\affiliation{Precursory Research for Embryonic Science and Technology (PRESTO), Japan Science and Technology Agency (JST), Kawaguchi 332-0012, Japan}
\affiliation{Center for Materials Research by Information Integration, National Institute for Materials Science (NIMS), Tsukuba 305-0047, Japan}

% \author{Atsushi Togo}
% \affiliation{Center for Elements Strategy Initiative for Structure Materials (ESISM), Kyoto University, Kyoto 606-8501, Japan}
% \affiliation{Center for Materials Research by Information Integration, National Institute for Materials Science (NIMS), Tsukuba 305-0047, Japan}

\author{J\"{o}rg Neugebauer}
\affiliation{Max-Planck-Institut f\"{u}r Eisenforschung GmbH, D-40237 D\"{u}sseldorf, Germany}

\author{Isao Tanaka}
% \email{tanaka@cms.mtl.kyoto-u.ac.jp}
\affiliation{Center for Elements Strategy Initiative for Structure Materials (ESISM), Kyoto University, Kyoto 606-8501, Japan}
\affiliation{Department of Materials Science and Engineering, Kyoto University, Kyoto 606-8501, Japan}
\affiliation{Center for Materials Research by Information Integration, National Institute for Materials Science (NIMS), Tsukuba 305-0047, Japan}
\affiliation{Nanostructures Research Laboratory, Japan Fine Ceramics Center, Nagoya 456-8587, Japan}

\date{\today}

\begin{abstract}
    A first-principles-based method for computing phonons of magnetic random solid solutions 
    including thermal magnetic fluctuations is developed.
    The method takes fluctuations of force constants (FCs) due to magnetic excitations
    as well as due to chemical disorder into account.
    The developed approach correctly predicts the experimentally observed unusual phonon hardening 
    of a transverse acoustic mode in Fe--Pd an Fe--Pt Invar alloys with increasing temperature. 
    This peculiar behavior, which cannot be explained within a conventional, harmonic picture, 
    turns out to be a consequence of thermal magnetic fluctuations.
\end{abstract}

% PACS, the Physics and Astronomy Classification Scheme.
% \pacs{
% 63.20.dk, % First-principles theory
% 71.15.Mb, % Density functional theory, local density approximation, gradient and other corrections
% 75.50.Bb, % Fe and its alloys
% 63.50.Gh, % Disordered crystalline alloys
% 63.20.-e  % Phonons in crystal lattices
% }

%\keywords{Suggested keywords}%Use showkeys class option if keyword
                              %display desired
\maketitle

%\tableofcontents

% %%%%%%%%%%%%%%%%%%%%%%%%%%%%%%%%%%%%%%%%
% \section{INTRODUCTION}
% \label{sec:introduction}
% %%%%%%%%%%%%%%%%%%%%%%%%%%%%%%%%%%%%%%%%

\begin{figure}[tbp]
\centering
\includegraphics[width=\linewidth]{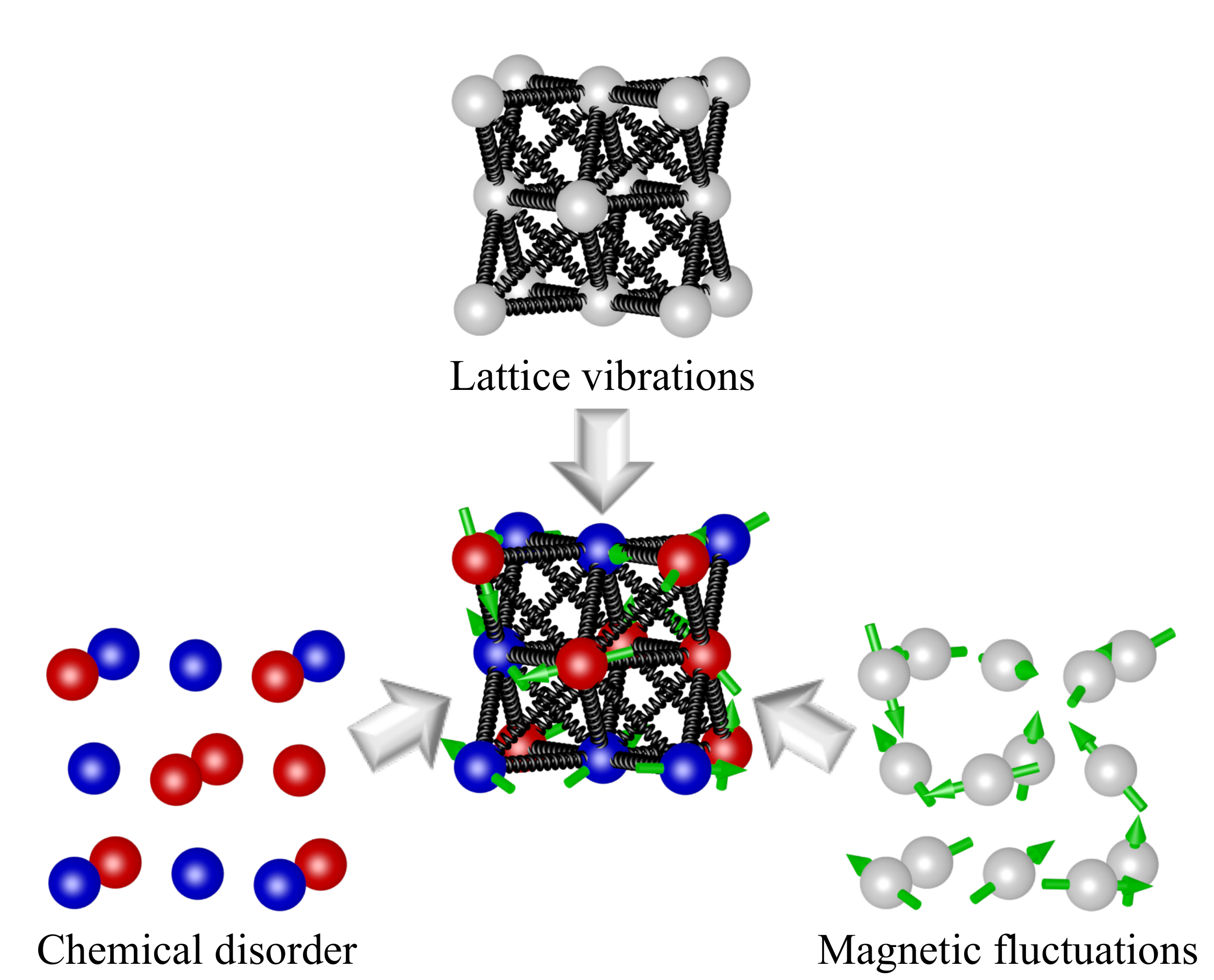}
\caption{
    (Color online)
    Sketch of the interplay between lattice vibrations, chemical disorder, and thermal magnetic fluctuations in a crystalline system.
    Spheres represent atoms, and their colors correspond to different chemical elements. 
    Coils connecting the spheres sketch lattice vibrations, 
    and green arrows represent magnetic moments on atoms.
    All three factors have to be taken simultaneously into account
    to simulate lattice vibrations of magnetic random solid solutions at finite temperature.
\label{fig:figure1}
}
\end{figure}

Magnetic random solid solutions represent a large and important class of crystalline materials 
ranging from structural materials such as steels
\cite{Munoz2011Positive, Razumovskiy2011, Vitos2002Elastic, Vitos2003},
including Invar alloys
\cite{Weiss1963, Schilfgaarde1999, Khmelevskyi2003Large, Yokoyama2011, Yokoyama2013},
up to multi-component magnetic high-entropy alloys
\cite{Zhang2015, Granberg2016Mechanism, Li2016}.
The simultaneous presence of chemical disorder and thermal magnetic fluctuations as well as their couplings to lattice vibrations play pivotal roles in many of these alloys.
Lattice vibrations largely dominate thermodynamic properties of materials
\cite{Fultz2010Vibrational}
and contribute to phase stability
\cite{Fultz2014Phase},
which is a key parameter for the computational design of new and innovative materials.
A computational scheme that can simulate lattice vibrations 
in magnetic random solid solutions by properly taking into account 
both, magnetic fluctuations as well as chemical disorder 
(as sketched in Fig.~\ref{fig:figure1}), is therefore of genuine importance. 

The delicate interactions between lattice vibrations,
chemical disorder,
and thermal magnetic fluctuations
can cause extreme and unusual physical properties.
A prominent example is the hardening of a transverse acoustic phonon mode and elastic constants with increasing temperature in Invar alloys \cite{Sato1982,Kastner1999}.
Since thermal expansion --- usually dominating the temperature dependence of phonon modes --- is in such alloys negligible 
\cite{Weiss1963, Schilfgaarde1999, Khmelevskyi2003Large, Yokoyama2011, Yokoyama2013}, 
the inclusion of explicit temperature dependent excitations, such as magnetic fluctuations, 
is critical to resolve such peculiarities.

In the last few years significant progress has been made for the computation of
lattice vibrations incorporating thermal magnetic fluctuations
for pure elements such as Fe
\cite{
Leonov2012, Leonov2014Electronic,  % DMFT
Ruban2012, *Ruban2014E,  % spin-wave
Kormann2012, Kormann2014, Ikeda2014Phonon, % SSA
Alling2016Strong}  % DLM-MD, Fe
and Ni
\cite{Kormann2016Impact}
as well as for several ordered magnetic compounds 
\cite{
Fennie2006,  %  Heisenberg-like model, ZnCr2O4
Shulumba2014,  % MD, CrN, May
Zhou2014, % SSA, CrN, Nov.
Gruner2015, % Random collinear arrangement, La--Fe--Si
Dutta2016}.  % Fixed-spin-moment approach, Ni2MnGa
While these methods clearly highlight the recent progress in first-principles-based thermodynamic approaches for magnetic systems,
they are still limited to chemically \textit{ordered} systems.
To investigate magnetic random solid solutions,
methods capable of including chemical disorder are required.
For random solid solutions \textit{without thermal magnetic fluctuations},
computational methods for lattice vibrations have also been advanced significantly in the last years.
Among such methods,
the itinerant coherent potential approximation (ICPA)
\cite{Ghosh2002, *Ghosh2003E}
and the band unfolding
\cite{Boykin2007, Allen2013, *Allen2013E, Ikeda2017unfolding}
have been used to successfully compute lattice vibrations for random solid solutions
\cite{
Dutta2009a,Dutta2009b,Dutta2010,Dutta2011,Granas2012,  % ICPA
Boykin2014, Huang2015, Zheng2016CMS, Overy2016, Ikeda2017unfolding}.  % Band-unfolding
However,
a computational method for lattice vibrations considering both,
thermal magnetic fluctuations and chemical disorder, is lacking so far.

We therefore propose a first-principles-based method to calculate lattice vibrations of magnetic random solid solutions, 
which addresses both, thermal magnetic fluctuations as well as chemical disorder on equal footing.
This is achieved by extending and combining above-mentioned disjunct approaches 
into a methodological framework,
which allows to predict temperature-dependent phonon spectra of magnetic random solid solutions.
To achieve this goal, we apply a two-step procedure and adiabatically decouple the (fast) magnetic and the (slow) chemical degrees of freedom.
To include thermal magnetic fluctuations,
we utilize force constants (FCs) which implicitly depend on the \textit{magnetic} temperature
\cite{Kormann2014}.
Such FCs 
are obtained using the spin-space averaging (SSA) method
\cite{Kormann2012}
in combination with quantum Monte Carlo (QMC) simulations for an effective Heisenberg spin Hamiltonian.
To account for chemical disorder, which induces variations of atomic masses and FCs among atomic sites,
we employ the ICPA
\cite{Ghosh2002, *Ghosh2003E}
and the band unfolding
\cite{Boykin2007, Allen2013, *Allen2013E, Ikeda2017unfolding}.

% %%%%%%%%%%%%%%%%%%%%%%%%%%%%%%%%%%%%%%%%
% \section{COMPUTATIONAL DETAILS}
% \label{sec:computational_details}
% %%%%%%%%%%%%%%%%%%%%%%%%%%%%%%%%%%%%%%%%

In the first step,
the FCs as functions of magnetic temperature are obtained as follows.
First the FCs in the low-temperature ``ideal'' ferromagnetic (FM) state,
where all the magnetic moments on atoms point to the same direction,
and the FCs in the high-temperature ``ideal'' paramagnetic (PM) state,
where the direction of the magnetic moments are fully disordered,
are calculated.	
Particularly the FCs in the ideal PM state are obtained using the SSA method
\cite{Kormann2012, Zhou2014},
i.e., by a statistical average over a large set of randomly distributed collinear magnetic moments.
Then the element-resolved FCs $\Phi_\textrm{M--M$'$} (T)$
for each pair of chemical elements M and M$'$
at temperature $T$ are obtained by extending the recently 
developed magnon-phonon coupling formalism 
\cite{Kormann2014} to random solid solutions as
%%%%%%%%%%
\begin{align}
    \Phi_{\textrm{M--M$'$}} (T)
    &=
    \alpha (T)
    \Phi^{\textrm{FM}}_{\textrm{M--M$'$}}
    +
    [1 - \alpha (T)]
    \Phi^{\textrm{PM}}_{\textrm{M--M$'$}},
    \label{eq:td_force_constants}
\end{align}
%%%%%%%%%%
where 
$\Phi^{\textrm{FM}}_{\textrm{M--M$'$}}$ and $\Phi^{\textrm{PM}}_{\textrm{M--M$'$}}$
denote the element-resolved FCs in the ideal FM and in the ideal PM states, respectively.
The interpolation parameter $\alpha (T)$ is directly related to the magnetic energy 
\cite{Kormann2014}
[see also Sec.~B in Supplemental Material (SM)].
Spin quantization effects for the magnetic energy and hence for $\alpha (T)$, 
being critical below $T_C$
\cite{Kormann2010},
are incorporated by performing numerically exact QMC simulations 
for an effective nearest-neighbor Heisenberg spin Hamiltonian
\cite{Kormann2010,Kormann2011,Kormann2013,Kormann2014}.
Here the solid solution is modeled by randomly distributing the chemical components 
having different spin values onto the magnetic sites.

In the second step, the thus obtained $\Phi_{\textrm{M--M$'$}} (T)$ is employed 
to derive the phonon spectrum of the magnetic random solid solution at any given $T$.
In principle both, the ICPA \cite{Ghosh2002, *Ghosh2003E} and the band unfolding 
\cite{Boykin2007, Allen2013, *Allen2013E, Ikeda2017unfolding}
can address the variations of atomic masses and FCs among atomic sites due to chemical disorder.
The ICPA has the advantage that it analytically incorporates the variations of atomic masses and FCs;
a \textit{fully} random solid solution is rigorously modeled in the ICPA.
This method, however, requires the numerical solution of a relatively complex set of equations
\cite{Ghosh2002, *Ghosh2003E},
making its extension to multicomponent systems challenging.
In contrast,
the band unfolding can be straightforwardly extended to such systems but is limited by the size of the supercell model;
undesired periodicity due to the limited size of the model may cause spurious features 
in the computed phonon spectra.
The possible influences of a limited supercell size can be, however,
straightforwardly eliminated by constructing a much larger, \textit{effective} supercell model 
in which the $\Phi_\textrm{M--M$'$} (T)$ computed from the original (smaller) supercell model are assigned
to the atomic sites (details of the mapping are given in the SM).
Such an effective supercell model has longer periodicity than the original supercell model
and therefore includes a larger number of distinct local configurations of chemical components.

We apply the developed approach to two experimentally well-studied magnetic alloys, 
namely to disordered face-centered cubic (fcc) Fe$_{0.72}$Pd$_{0.28}$ and Fe$_{0.72}$Pt$_{0.28}$ alloys. 
Both Invar alloys reveal the aforementioned characteristic phonon hardening of a $\langle 110 \rangle$ (in fractional coordinates for the conventional fcc unit cell) transverse acoustic mode 
when heated up above the Curie temperature, $T_{C}$
\cite{Sato1982, Kawald1990}.
In the following we focus on
(i) the impact of chemical disorder as well as the performance of 
the ICPA and the band unfolding to incorporate it, and, in particular, 
(ii) the impact of thermal magnetic fluctuations on the phonon spectra.

Chemical and magnetic (for the ideal PM state) disorder in these alloys
were simulated by special quasirandom structures (SQSs)
\cite{Zunger1990}.
The SQSs in this study were constructed on the 32-atom
$2 \times 2 \times 2$ supercell of the conventional fcc unit cell
(see also Sec.~A in SM).
Chemical compositions for the SQS supercell models were chosen to be
Fe$_{0.75}$Pd$_{0.25}$ and Fe$_{0.75}$Pt$_{0.25}$,
being close to the experimental ones.
$\Phi^{\textrm{FM}}_\textrm{M--M$'$}$ and
$\Phi^{\textrm{PM}}_\textrm{M--M$'$}$ 
were calculated for the SQS supercell models  
using the finite-displacement method.
% finite atomic displacements of 0.01~{\AA}.
Among the computed FCs, those up to the fourth nearest neighbors were taken into account in the following steps.
$\alpha (T)$ in Eq.~\eqref{eq:td_force_constants} was determined 
using the effective Heisenberg spin Hamiltonian for a $7 \times 7 \times 7$ supercell including 1372 magnetic sites 
(see also Sec.~B in SM).
In the band-unfolding procedure,
effective 864-atom $6 \times 6 \times 6$ supercell models,
where $\Phi_{\textrm{M--M$'$}} (T)$ obtained from
the $2 \times 2 \times 2$-SQS models were assigned among the atomic sites,
were employed.
Since thermal expansion is small for the chosen Invar alloys, fixed lattice constants were applied for the computer simulations,
which were taken from experimental data at room temperature;
3.755 {\AA} \cite{Sato1982} (disordered fcc Fe$_{0.72}$Pd$_{0.28}$) and
3.749 {\AA} \cite{Sumiyama1976} (disordered fcc Fe$_{0.72}$Pt$_{0.28}$).
Electronic structures were calculated in the framework of density-functional theory 
within the generalized gradient approximation of the Perdew-Burke-Ernzerhof from
\cite{Perdew1996}  % J. P. Perdew, K. Burke, and M. Ernzerhof, Phys. Rev. Lett. 77, 3865 (1996)
using the plane-wave basis projector augmented wave method
\cite{Blochl1994}  % P. E. Bl\”{o}chl, Phys. Rev. B 50, 17953 (1994)
as implemented in the \textsc{vasp} code
\cite{Kresse1995, Kresse1996, Kresse1999}.
% G. Kresse, J. non-Cryst. Solids 192-193, 222 (1995),
% G. Kresse and J. Furthmüller, Comput. Mater. Sci. 6, 15 (1996),
% G. Kresse and D. Joubert, Phys. Rev. B 59, 1758 (1999),
%
A plane-wave energy cutoff of 350~eV was used.
Internal atomic positions were fully relaxed while keeping the lattices of the
SQS models fixed.

% %%%%%%%%%%%%%%%%%%%%%%%%%%%%%%%%%%%%%%%%
% \section{RESULTS AND DISCUSSION}
% \label{sec:results}
% %%%%%%%%%%%%%%%%%%%%%%%%%%%%%%%%%%%%%%%%

\begin{figure}[tbp]
\centering
\includegraphics[width=\linewidth]{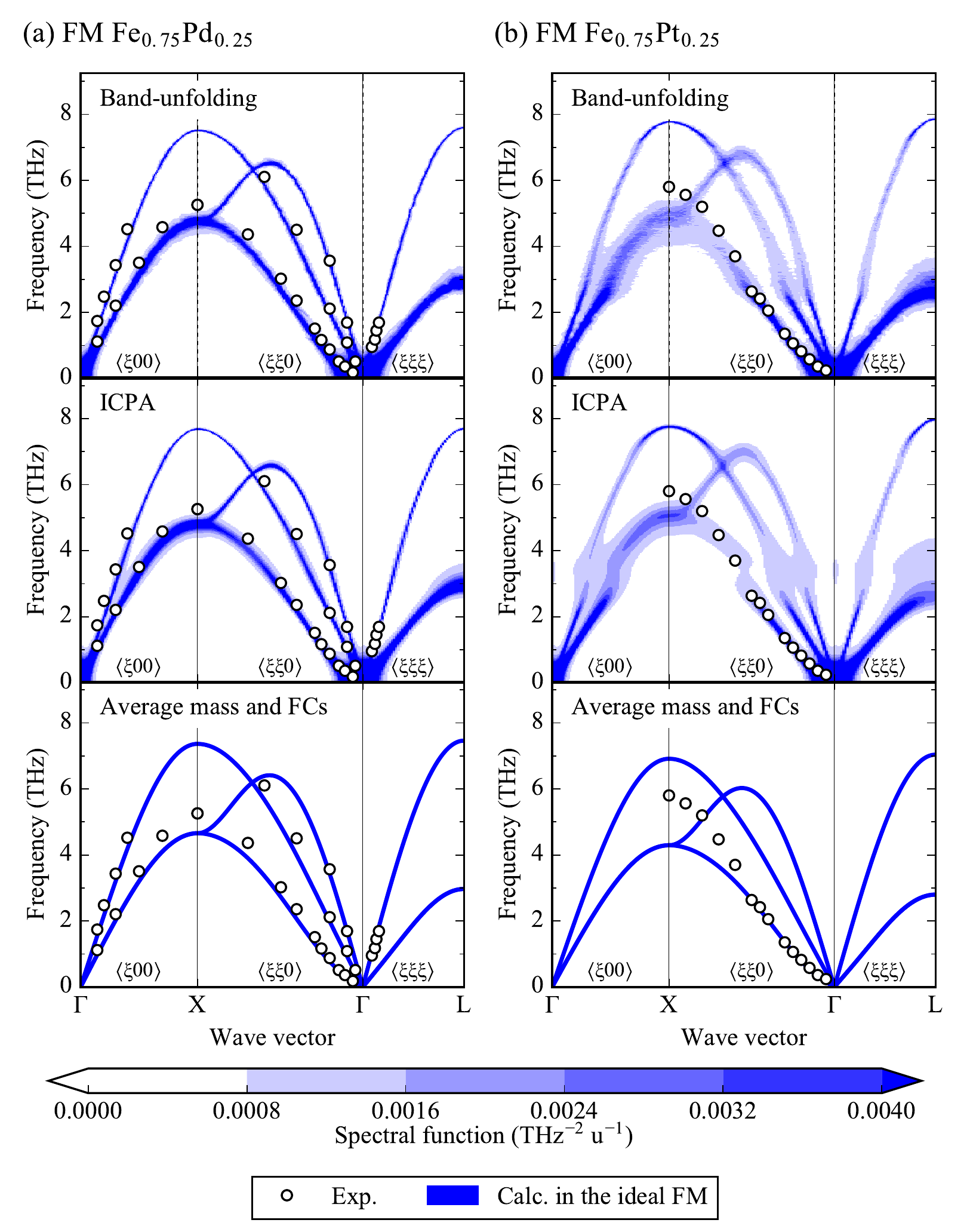}
\caption{
(Color online)
Phonon spectra of chemically disordered fcc (a) Fe$_{0.75}$Pd$_{0.25}$ 
and (b) Fe$_{0.75}$Pt$_{0.25}$
in the ideal FM state calculated using the band unfolding (top panels)
and the ICPA (middle panels) methods.
Blue color-contour corresponds to the magnitude of the spectral functions. 
Bottom panels show the phonon dispersion relations 
calculated using the average atomic masses and FCs.
White circles show the experimental phonon frequencies 
for disordered fcc Fe$_{0.72}$Pd$_{0.28}$
\cite{Sato1982}
and Fe$_{0.72}$Pt$_{0.28}$
\cite{Kastner1999}
at room temperature,
where these disordered alloys are in the FM phase.
\label{fig:phonons_FM}
}
\end{figure}

In order to separately discuss the impact of chemical disorder and thermal magnetic fluctuations,
we first focus on the ideal FM state [$\alpha(T = 0) = 1$], i.e.,
in the absence of thermal magnetic fluctuations.
Figure~\ref{fig:phonons_FM} shows the phonon spectra
of Fe$_{0.75}$Pd$_{0.25}$ and Fe$_{0.75}$Pt$_{0.25}$
in the ideal FM state
calculated using the band unfolding (top panels) and the ICPA (middle panels).
%%%%%%%%%%%%%%%%%%%%%%%%%%%%%%%%%%%%%%%%
% 1. Comparison between the two methods
%%%%%%%%%%%%%%%%%%%%%%%%%%%%%%%%%%%%%%%%
The results obtained from the two different methods are very similar to each other for both alloys,
although the formalisms of the ICPA and of the band unfolding look very different.
From this analysis we conclude that both methods incorporate 
the impact of chemical disorder into the calculations of phonon spectra in a similar quality.
%%%%%%%%%%%%%%%%%%%%%%%%%%%%%%%%%%%%%%%%
% 2. Experimental results
%%%%%%%%%%%%%%%%%%%%%%%%%%%%%%%%%%%%%%%%
It is also found that the peak positions of the computed phonon spectra are in good agreement with experimental phonon frequencies.

%%%%%%%%%%%%%%%%%%%%%%%%%%%%%%%%%%%%%%%%
% 3. Chemical disorder: Two methods
% Linewidths
%%%%%%%%%%%%%%%%%%%%%%%%%%%%%%%%%%%%%%%%
The spectra also show phonon broadening originating from the variations of atomic masses and FCs
among atomic sites due to the chemical disorder.
The broadening in the Fe$_{0.75}$Pd$_{0.25}$ is relatively small,
while Fe$_{0.75}$Pt$_{0.25}$ shows large broadening in the frequency region around 3--5~THz.
The increased broadening in Fe$_{0.75}$Pt$_{0.25}$ occurs probably because of
the large differences of atomic masses and FCs among the chemical components.
The atomic mass of Pt relative to Fe ($\approx 3.5$) is much larger than that of Pd ($\approx 1.9$).
The FCs in Fe$_{0.75}$Pt$_{0.25}$ are also largely different 
among the distinct combinations of the chemical elements 
compared with those in Fe$_{0.75}$Pd$_{0.25}$
(see also Sec.~D in SM).
% $195.084 / 55.845 \simeq  3.5$
% $106.42 / 55.845 \simeq  1.9$

To elucidate the impacts of the variations of atomic masses and FCs among atomic
sites more clearly,
the phonon dispersion relations are also calculated in the \textit{absence} of FC and mass fluctuations
using the concentration-weighted average atomic mass, $\bar{m}$, and
the \textit{crystallographically-symmetrized} FCs $\bar{\Phi}$.
$\bar{\Phi}$ is obtained by first applying 
each symmetry operation of the fcc structure to the original FCs 
and then taking the average of the transformed FCs \textit{irrespective of} chemical components.
The results are shown in the bottom panels in Fig.~\ref{fig:phonons_FM}.
Note that by construction, no phonon broadening is obtained in this case.
The phonon frequencies derived from $\bar{m}$ and $\bar{\Phi}$
agree reasonably well with experiment for Fe$_{0.75}$Pd$_{0.25}$.
For Fe$_{0.75}$Pt$_{0.25}$ the deviations are, however, significant, in particular around the X point.
This indicates the importance of taking the variations of atomic masses and FCs among atomic sites into account
for accurate phonon computations of random solid solutions.

Having verified the importance of an appropriate treatment of chemical disorder
in random solid solutions,
we next analyze the impact of thermal magnetic fluctuations on the phonon spectra
from the viewpoint of their temperature dependence.
The temperature-dependent spectra are obtained using 
$\Phi_{\textrm{M--M$'$}} (T)$ from Eq.~\eqref{eq:td_force_constants}.
Calculations are carried out at three representative characteristic temperatures, 
i.e., below, near, and above $T_C$
($T_{C}$ is 575~K for Fe$_{0.72}$Pd$_{0.28}$ \cite{Sato1982}
and 367~K for Fe$_{0.72}$Pt$_{0.28}$ \cite{Kawald1990}).
We focus in particular on the lowest-frequency phonon branch along the $\langle 110 \rangle$ direction,
which shows the unusual hardening with increasing temperature. 
Note that within the harmonic approximation and in absence of explicit temperature dependent excitations,
phonon spectra only implicitly depend on temperature via the volume expansion. 
However, as mentioned above, the considered Invar alloys reveal negligible expansion below $T_{C}$. 
A {\it conventional} harmonic approximation would therefore predict a temperature-independent (i.e. constant) phonon spectra 
and cannot explain the experimental data.

\begin{figure}[tbp]
\centering
\includegraphics[width=\linewidth]{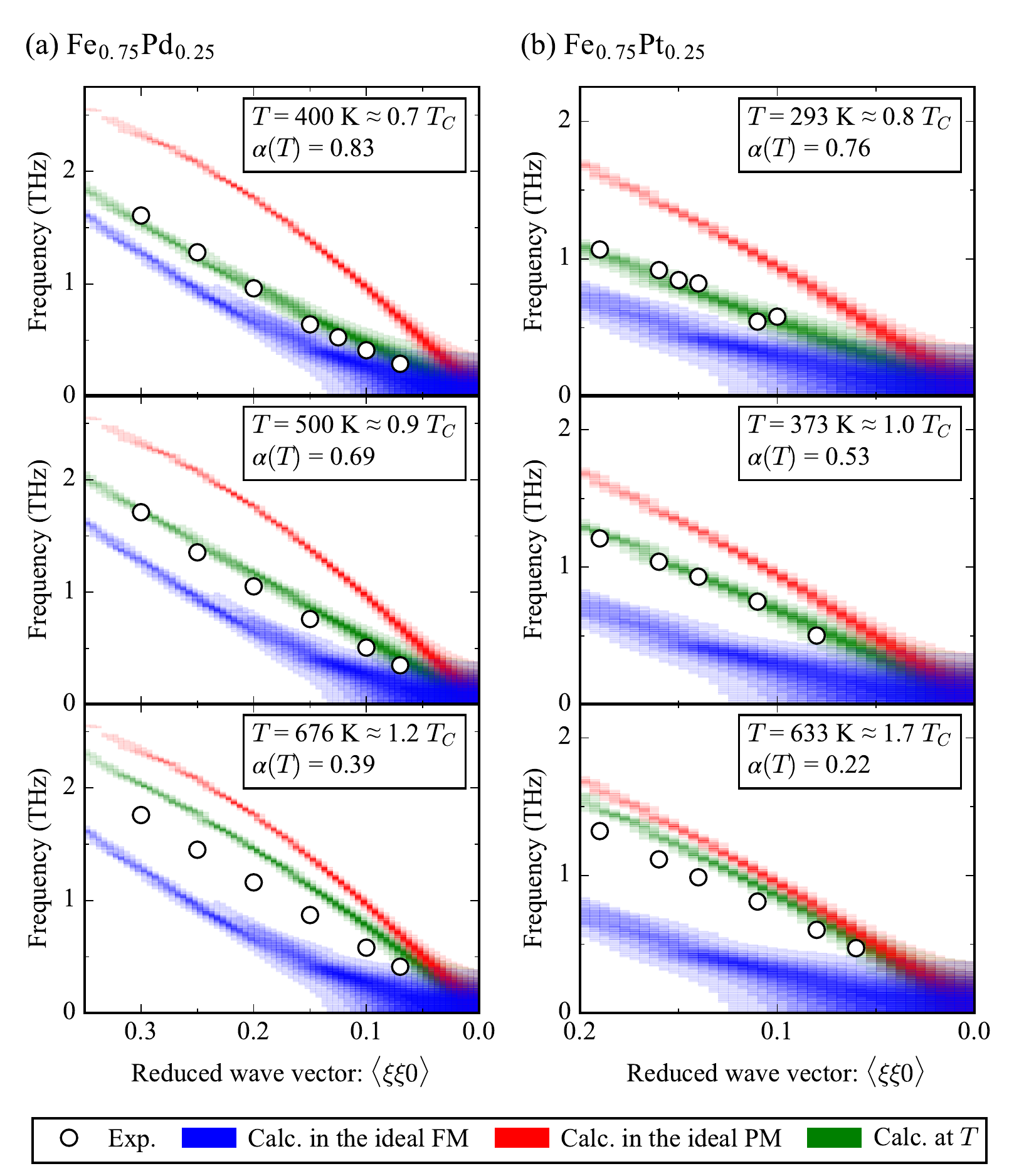}
\caption{
    (Color online)
    Temperature dependence of the phonon spectra for the lowest-frequency branch 
    along the $\langle 110 \rangle$ direction
    (in fractional coordinates for the conventional fcc unit cell)
    calculated using the band unfolding.
    (a) Fe$_{0.75}$Pd$_{0.25}$. 
    (b) Fe$_{0.75}$Pt$_{0.25}$.
    Blue, red, and green color-contours show the results
    obtained from the FCs in the ideal FM, in the ideal PM, and at the corresponding
    temperatures obtained from Eq.~(\ref{eq:td_force_constants}), respectively.
    $\alpha (T)$ used to obtain the FCs at the corresponding temperatures are
    also given in the panels.
    White circles show the experimental phonon frequencies for
    Fe$_{0.72}$Pd$_{0.28}$
    \cite{Sato1982}
    and
    Fe$_{0.72}$Pt$_{0.28}$
    \cite{Endoh1979}.
    \label{fig:phonons_temperature}
}
\end{figure}

The results employing the band-unfolding method are shown in Figure~\ref{fig:phonons_temperature}
\footnote{Note that similarly to the discussed ideal FM state (Fig.~\ref{fig:phonons_FM}), 
the ICPA results are again in excellent agreement with the band-unfolding results
for
% (the ideal PM state and)
the intermediate temperatures and therefore shown
only in SM.}.
For comparison,
the results for the ideal FM and the ideal PM states are also shown.
The peak positions of the computed temperature-dependent phonon spectra 
are in agreement with experimental phonon frequencies
at the corresponding temperatures,
while the results for the ideal FM and the ideal PM states provide 
the lower and the upper bounds, respectively.
This indicates that thermal magnetic fluctuations are crucial to reproduce the experimental data.

%
% 1. Comparison between the FM and the PM limits (Invar effects)
%

Our simulations further indicate the presence of considerable magnetic short-range ordering (SRO) 
(i.e., a finite value of $\alpha(T)$ in Fig.~\ref{fig:phonons_temperature}) even above $T_{C}$, 
similar as already observed for pure Fe \cite{Kormann2014}. 
Without taking magnetic SRO into account, i.e., by limiting the simulations solely to the ideal PM state, 
the agreement with experiment is lost.  
Magnetic SRO hence plays a crucial role, not only for pure Fe, but also for Fe-based alloys. 
We also observe slight deviations between our simulations and the experiment above $T_{C}$. 
We attribute these deviations to the increase of lattice constants 
with temperature above $T_{C}$ (the Invar effect is lost),
which is not yet included in our calculations.

%
% 3. Martensitic transformations
%
From the simulated temperature dependence of the phonon modes,
we can also deduce information on the martensitic transformation for the investigated alloys.
In Fig.~\ref{fig:phonons_temperature},
the slope of the lowest-frequency branch along the $\langle 110 \rangle$ direction 
around the $\Gamma$ point largely depends on temperature 
for both Fe$_{0.75}$Pd$_{0.25}$ and Fe$_{0.75}$Pt$_{0.25}$;
the slope increases with increasing temperature. 
The slope of this branch is associated with the elastic stiffness constant
$C' \equiv (C_{11} - C_{12}) / 2$,
which is related to the dynamical stability for the martensitic transformation 
along the Bain path
\cite{Bain1924}.
The softening of this branch when lowering the temperature is considered as a precursor of the martensitic transformation.
In experiments disordered fcc Fe$_{1-x}$Pd$_{x}$ and Fe$_{1-x}$Pt$_{x}$
($x =$ 0.25--0.33) indeed show a martensitic transformation;
for example, Fe$_{0.72}$Pd$_{0.28}$ and Fe$_{0.73}$Pt$_{0.27}$ show a martensitic transformation 
at 314.5~K \cite{Arabi-Hashemi2015} and approximately at 125~K \cite{Sumiyama1983Characteristic}, respectively.
Thus, our computed result suggests that
the martensitic transformation for these alloys originates from magnon-phonon coupling.

% %%%%%%%%%%%%%%%%%%%%%%%%%%%%%%%%%%%%%%%%
% \section{CONCLUSIONS}
% \label{sec:conclusions}
% %%%%%%%%%%%%%%%%%%%%%%%%%%%%%%%%%%%%%%%%

In summary,
we propose a first-principles-based method to incorporate 
both, thermal magnetic fluctuations and chemical disorder
into a unified computational framework of phonon calculations for magnetic random solid solutions.
Chemical disorder, 
which leads to variations of atomic masses and FCs among atomic sites,
is taken into account using the ICPA and the band unfolding in combination with SQSs.
Thermal magnetic fluctuations are incorporated using the SSA 
in combination with QMC simulations for an effective Heisenberg spin Hamiltonian.

The proposed methodology is applied to Fe--Pd and Fe--Pt Invar alloys.
Both, the ICPA and the band unfolding are found to be equally capable 
for computing phonon spectra of the chemically disordered alloys.
Both methods also reveal phonon broadening due to the chemical disorder.
Taking thermal magnetic fluctuations into account,
the developed approach shows excellent agreement between 
the temperature-dependent peak positions of the computed phonon spectra and experimental phonon frequencies.
In particular, the approach correctly reproduces the experimentally observed 
unusual hardening in iron-based Invar alloys at high temperatures and shows that magnetic fluctuations are responsible for it. 
Finite-temperature magnon-phonon contributions also cause the experimentally observed softening 
of the elastic constant $C'$
and hence trigger the martensitic transformation in these alloys.

The proposed approach is general and can be straightforwardly employed to examine 
other complex magnetic random solid solutions such as magnetic high-entropy alloys,
which are expected to reveal complex physical mechanisms caused by 
the interplay between thermal magnetic fluctuations, chemical disorder, and lattice vibrations.

\begin{acknowledgments}
Funding by the Ministry of Education, Culture, Sports,
Science, and Technology (MEXT), Japan, through Elements Strategy Initiative for
Structural Materials (ESISM) of Kyoto University,
by the Japan Society for the Promotion of Science (JSPS) KAKENHI 
Grant-in-Aid for Young Scientist (B) (Grant No.~16K18228),
% From Fritz
by the European Research Council under the EU's 7th Framework Programme
(FP7/2007-2013)/ERC Grant agreement 290998,
by the Deutsche Forschungsgemeinschaft (DFG) for the scholarship KO 5080/1-1,
and by the DFG for their funding within the priority programme SPP 1599
are gratefully acknowledged.
\end{acknowledgments}

% \bibliography{main.bib,DFT.bib,Cr.bib,Fe.bib,Ni.bib,CrN.bib,sqs.bib,ICPA.bib,unfolding.bib,FeNi.bib,HEAs.bib,ikeda.bib}
%merlin.mbs apsrev4-1.bst 2010-07-25 4.21a (PWD, AO, DPC) hacked
%Control: key (0)
%Control: author (72) initials jnrlst
%Control: editor formatted (1) identically to author
%Control: production of article title (-1) disabled
%Control: page (0) single
%Control: year (1) truncated
%Control: production of eprint (0) enabled
%

\clearpage


\section*{Supplementary material}

# Supplemental Material: Temperature-dependent phonon spectra of magnetic random solid solutions

Yuji Ikeda,[1,*] Fritz Körmann,[2,3] Biswanath Dutta,[3] Abel Carreras,[4] Atsuto Seko,[1,4,5,6] Jörg Neugebauer,[3] and Isao Tanaka[1,4,6,7]

[1]*Center for Elements Strategy Initiative for Structure Materials (ESISM), Kyoto University, Kyoto 606-8501, Japan*
[2]*Department of Materials Science and Engineering, Delft University of Technology, Mekelweg 2, 2628 CD Delft, Netherlands*
[3]*Max-Planck-Institut für Eisenforschung GmbH, D-40237 Düsseldorf, Germany*
[4]*Department of Materials Science and Engineering, Kyoto University, Kyoto 606-8501, Japan*
[5]*Precursory Research for Embryonic Science and Technology (PRESTO), Japan Science and Technology Agency (JST), Kawaguchi 332-0012, Japan*
[6]*Center for Materials Research by Information Integration, National Institute for Materials Science (NIMS), Tsukuba 305-0047, Japan*
[7]*Nanostructures Research Laboratory, Japan Fine Ceramics Center, Nagoya 456-8587, Japan*
(Dated: February 8, 2017)

## A. Generation of special quasirandom structures

Here the details of the special quasirandom structures (SQSs) [S1] employed to calculate the force constants (FCs) in $\mathrm{Fe}_{0.75}\mathrm{Pd}_{0.25}$ and $\mathrm{Fe}_{0.75}\mathrm{Pt}_{0.25}$ are described. The SQSs were constructed for the 32-atom $2 \times 2 \times 2$ supercell of the conventional face-centered cubic (fcc) unit cell.

The ideal ferromagnetic (FM) state was modeled using a binary SQS with the composition of $\mathrm{A}_{0.75}\mathrm{B}_{0.25}$. "A" stands for Fe with spin-up magnetic moments, and "B" stands for Pd or Pt. A variable $\sigma_i$ is given as $+1$ or $-1$ when the $i$th atomic site is occupied by A or B, respectively. The basis of the correlation functions $\phi(\sigma_i)$ is given as $\phi(\sigma_i) = \sigma_i$. Table S1 summarizes the values of the correlation functions for the binary SQS, where the employed triplet and quartet clusters are sketched in Fig. S1.

The ideal paramagnetic (PM) state was modeled using a ternary SQS with the composition of $\mathrm{A}_{0.375}\mathrm{B}_{0.375}\mathrm{C}_{0.25}$. "A" and "B" stand for magnetically disordered Fe atoms with spin-up and spin-down magnetic moments, respectively, while "C" stands for Pd or Pt atoms. Note that due to the negligibly small magnetic moments on Pd and Pt in the ideal PM state ($< 0.1\,\mu_B$), the randomization of magnetic moments was omitted for these elements. $\sigma_i$ is given as $+1$, $-1$, or $0$ when the $i$th atomic site is occupied by A, B, or C, respectively. The basis of the correlation functions is $\phi_1(\sigma_i) = \sqrt{3/2}\,\sigma_i$ and $\phi_2(\sigma_i) = -\sqrt{2}(1 - 3/2\,\sigma_i)$. Table S2 summarizes the values of the pair correlation functions for the ternary SQS.

The SQSs were obtained by simulated annealing [S2, S3] as implemented in the CLUPAN code [S4, S5].

---

[*] ikeda.yuji.6m@kyoto-u.ac.jp

TABLE S1. Values of the correlation functions for the binary SQS with the composition of $\mathrm{A}_{0.75}\mathrm{B}_{0.25}$ employed in this study. $n$NN is the abbreviation of "$n$th nearest neighbor".

|         |            | Correlation function |
|---------|------------|----------------------|
| Pair    | 1NN        | 1/4                  |
|         | 2NN        | 1/4                  |
|         | 3NN        | 3/16                 |
|         | 4NN        | 1/4                  |
|         | Random     | 1/4                  |
| Triplet | Fig. S1(a) | 1/8                  |
|         | Random     | 1/8                  |
| Quartet | Fig. S1(b) | 1/16                 |
|         | Random     | 0                    |

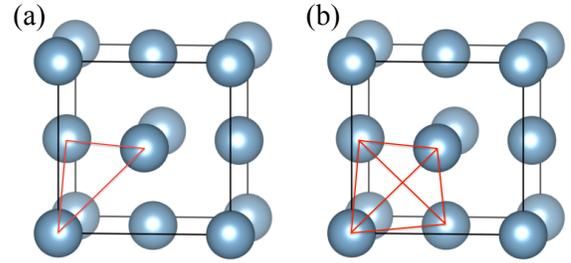

FIG. S1. (a) Triplet and (b) quartet clusters whose correlation functions are shown in Table S1. Spheres represent atoms in the fcc structure.

## B. Magnetic energy and mixing parameter $\alpha(T)$ for the FCs at intermediate temperatures

The mixing parameter $\alpha(T)$ in Eq. (1) in the main text is related to the internal magnetic energy, $E_\mathrm{mag}(T)$, as [S6]

$$\alpha(T) = \frac{E_\mathrm{mag}(T) - E_\mathrm{mag}^\mathrm{PM}}{E_\mathrm{mag}^\mathrm{FM} - E_\mathrm{mag}^\mathrm{PM}}, \quad (\mathrm{S1})$$

where $E_\mathrm{mag}^\mathrm{FM} \equiv E_\mathrm{mag}(T = 0)$ and $E_\mathrm{mag}^\mathrm{PM} \equiv E_\mathrm{mag}(T \to \infty)$ represent the internal magnetic energies of the low-temperature ideal FM state and the high-temperature ideal PM state, respectively. The internal magnetic energies were obtained

2TABLE S2. Values of the pair correlation functions for the ternary SQS with the composition of $A_{0.375}B_{0.375}C_{0.25}$ employed in this study. $n$NN is the abbreviation of "$n$th nearest neighbor".

|  | Pair correlation function | | |
| --- | --- | --- | --- |
|  | $\langle\phi_1(\sigma_i)\phi_1(\sigma_j)\rangle$ | $\langle\phi_1(\sigma_i)\phi_2(\sigma_j)\rangle$ | $\langle\phi_2(\sigma_i)\phi_2(\sigma_j)\rangle$ |
| 1NN | 0 | 0 | 1/32 |
| 2NN | $-3/32$ | 0 | 1/32 |
| 3NN | $-1/128$ | 0 | $-5/128$ |
| 4NN | 1/32 | 0 | 1/32 |
| Random | 0 | 0 | 1/32 |

TABLE S3. Average magnetic moments on the chemical components extracted from the $2\times 2\times 2$ supercell models.

|  | Magnetic moment ($\mu_B$) | | |
| --- | --- | --- | --- |
|  | Fe | Pd | Pt |
| $Fe_{0.75}Pd_{0.25}$ in the ideal FM | 2.77 | 0.27 |  |
| $Fe_{0.75}Pd_{0.25}$ in the ideal PM | 2.57 | 0.03 |  |
| $Fe_{0.75}Pt_{0.25}$ in the ideal FM | 2.75 |  | 0.32 |
| $Fe_{0.75}Pt_{0.25}$ in the ideal PM | 2.53 |  | 0.06 |

by mapping the magnetic system onto an effective nearest-neighbor Heisenberg spin Hamiltonian [S7–S10]

$$\mathcal{H}^{\mathrm{mag}} = -\sum_{\langle i,j\rangle} J_{ij}\hat{\mathbf{S}}_i \cdot \hat{\mathbf{S}}_j, \tag{S2}$$

where $\hat{\mathbf{S}}_i$ is the spin operator at the $i$th site, and $J_{ij}$ represents an effective magnetic interaction between the $i$th and the $j$th magnetic sites.

The Hamiltonian in Eq. (S2) was solved using quantum Monte Carlo (QMC) calculations based on the direct loop algorithm [S11] in the stochastic series-expansion technique as implemented in the ALPS code [S12, S13]. The QMC calculations involved $1.5\times 10^7$ steps including thermalization and statistical averaging. The disordered alloys were modeled by randomly distributing the chemical components onto the 1372 sites of the $7\times 7\times 7$ supercell based on the conventional fcc unit cell.

The spin quantum number $S_i$, which satisfies $\mathbf{S}_i^2 = S_i(S_i+1)$, was extracted from the first-principles calculations for the $2\times 2\times 2$-supercell models via $M_i = g\mu_B S_i \approx 2\mu_B S_i$, where $M_i$ denotes the magnetic moment, $\mu_B$ is the Bohr magneton, and $g\approx 2$ the Landé factor [S10]. $M_i$ was calculated for each chemical component as the average over the magnetic moments of the corresponding chemical component in the supercell models, which is summarized in Table S3. The differences of the Fe magnetic moments in the ideal FM and the ideal PM states are approximately 0.2 $\mu_B$. This reveals only minor contributions from longitudinal magnetic fluctuations, which are neglected in the present study. Due to the quantum nature of Eq. (S2), only multiples of 1/2 for $S_i$ are accessible. The result for non-half integer spin quantum numbers were therefore mimicked by interpolating the results for $S_i = 1$ and $S_i = 3/2$, as in detail described in [S10]. The average of the Fe magnetic moments in the two ideal magnetic states were adopted for the interpolated values while Pd and Pt were treated as vacant sites due to their negligible magnetic moments.

The effective $J_{ij}$ was determined so that the Hamiltonian in Eq. (S2) reproduces the experimental Curie temperature (575 K and 367 K for $Fe_{0.72}Pd_{0.28}$ and $Fe_{0.72}Pt_{0.28}$, respectively [S14, S15]), as in detail described in [S9, S10]. Given the vanishing spin moments on the Pd and Pt sites at elevated temperatures (see above), the effective nearest-neighbor interaction is assumed to be non-zero only between nearest-neighbor Fe magnetic spins.

### C. Phonon spectral function in the band unfolding

Typically in the band unfolding, the phonon spectral function $A(\mathbf{k},\omega)$ at the wave vector $\mathbf{k}$ and at the frequency $\omega$ is given as [S16]

$$A(\mathbf{k},\omega) = \sum_J \left|[\hat{P}^{\mathbf{k}}\tilde{\mathbf{v}}(\mathbf{K},J)]_l\right|^2 \delta[\omega - \omega(\mathbf{K},J)], \tag{S3}$$

where $\hat{P}^{\mathbf{k}}$ is the projection operator for $\mathbf{k}$, $\tilde{\mathbf{v}}(\mathbf{K},J)$ is the $J$th "phase-weighted" phonon-mode eigenvector of a supercell model at the wave vector $\mathbf{K}$, $\omega(\mathbf{K},J)$ is the corresponding phonon frequency, and the subscript $l$ is the index for unit cells. The thus obtained $A(\mathbf{k},\omega)$, however, does not satisfy the same normalization condition as that defined for the ICPA [S17]. To apply the same normalization condition, the definition of $A(\mathbf{k},\omega)$ is slightly modified in this study as

$$A(\mathbf{k},\omega^2) = \left(\sum_X \frac{c_X}{m_X}\right)\sum_J \left|[\hat{P}^{\mathbf{k}}\tilde{\mathbf{v}}(\mathbf{K},J)]_l\right|^2 \delta\{\omega^2 - [\omega(\mathbf{K},J)]^2\}, \tag{S4}$$

where $m_X$ and $c_X$ are the atomic mass and composition ratio for the chemical element X, respectively. The differences from the original are that squared frequencies are employed instead of raw frequencies and that the constant $\sum_X c_X/m_X$ is multiplied.

### D. Distributions of FCs

Figures S2 and S3 show the distributions of the first-nearest neighbor (1NN) FCs with respect to interatomic distance for $Fe_{0.75}Pd_{0.25}$ and $Fe_{0.75}Pt_{0.25}$, respectively, in the ideal FM state. Figures S4 and S5 show the results in the ideal PM state. All the results are obtained from the $2\times 2\times 2$-SQS models. Tables S4 and S5 summarize the average and the standard deviation of the FCs.

The values of the element-resolved FCs are widely distributed in both alloys. The FC values tend to be ordered in magnitude from Pd–Pd/Pt–Pt over Fe–Pd/Fe–Pt to Fe–Fe pairs in $Fe_{0.75}Pd_{0.25}/Fe_{0.75}Pt_{0.25}$. The differences among the element-resolved FCs are larger in $Fe_{0.75}Pt_{0.25}$ than in $Fe_{0.75}Pd_{0.25}$. Specifically, the FCs of Fe–Fe pairs are similar in both alloys, while the FCs of Pt–Pt pairs in $Fe_{0.75}Pt_{0.25}$ tend to be larger in magnitude than the FCs of Pd–Pd pairs in $Fe_{0.75}Pd_{0.25}$.

TABLE S4. Average and standard deviation (SD) of the FCs as well as interatomic distances between the 1NN atomic pairs for $Fe_{0.75}Pd_{0.25}$.

| | | Interatomic distance (Å) | | FCs (eV/Å$^2$) | | | | | | | |
|---|---|---|---|---|---|---|---|---|---|---|---|
| | | | | $xx$ | | $xy$ | | $xz$ | | $zz$ | |
| | | Average | SD | Average | SD | Average | SD | Average | SD | Average | SD |
| Ideal FM | Fe–Fe | 2.640 | 0.057 | −0.810 | 0.162 | −0.945 | 0.197 | 0.000 | 0.106 | 0.161 | 0.114 |
| | Fe–Pd | 2.668 | 0.044 | −1.072 | 0.268 | −1.291 | 0.310 | 0.000 | 0.117 | 0.136 | 0.076 |
| | Pd–Pd | 2.736 | 0.040 | −1.424 | 0.341 | −1.571 | 0.342 | 0.000 | 0.090 | 0.153 | 0.062 |
| Ideal PM | Fe–Fe | 2.627 | 0.095 | −0.496 | 0.228 | −0.524 | 0.248 | 0.000 | 0.134 | −0.081 | 0.089 |
| | Fe–Pd | 2.692 | 0.066 | −0.836 | 0.302 | −0.952 | 0.357 | 0.000 | 0.095 | −0.001 | 0.065 |
| | Pd–Pd | 2.727 | 0.037 | −1.531 | 0.320 | −1.746 | 0.346 | 0.000 | 0.055 | 0.140 | 0.069 |

TABLE S5. The same as Table S4, but for $Fe_{0.75}Pt_{0.25}$.

| | | Interatomic distance (Å) | | FCs (eV/Å$^2$) | | | | | | | |
|---|---|---|---|---|---|---|---|---|---|---|---|
| | | | | $xx$ | | $xy$ | | $xz$ | | $zz$ | |
| | | Average | SD | Average | SD | Average | SD | Average | SD | Average | SD |
| Ideal FM | Fe–Fe | 2.639 | 0.044 | −0.881 | 0.135 | −1.053 | 0.177 | 0.000 | 0.093 | 0.177 | 0.097 |
| | Fe–Pt | 2.654 | 0.043 | −1.298 | 0.296 | −1.581 | 0.333 | 0.000 | 0.129 | 0.205 | 0.066 |
| | Pt–Pt | 2.753 | 0.048 | −1.689 | 0.534 | −1.886 | 0.510 | 0.000 | 0.117 | 0.318 | 0.088 |
| Ideal PM | Fe–Fe | 2.629 | 0.086 | −0.479 | 0.226 | −0.509 | 0.249 | 0.000 | 0.139 | −0.080 | 0.107 |
| | Fe–Pt | 2.675 | 0.063 | −0.981 | 0.367 | −1.121 | 0.412 | 0.000 | 0.118 | 0.032 | 0.085 |
| | Pt–Pt | 2.740 | 0.039 | −1.842 | 0.448 | −2.085 | 0.448 | 0.000 | 0.094 | 0.256 | 0.070 |

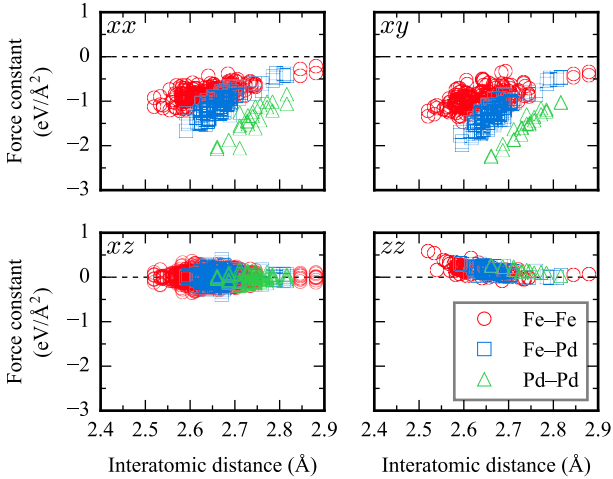

FIG. S2. Distributions of the 1NN FCs with respect to interatomic distance calculated using the $2 \times 2 \times 2$-SQS model for disordered fcc $Fe_{0.75}Pd_{0.25}$ in the ideal FM state. The atoms are assumed to be on $(0, 0, 0)$ and $(1/2, 1/2, 0)$ in fractional coordinates of the conventional fcc unit cell. Each panel corresponds to the symmetrically inequivalent element of the 1NN FCs specified at the upper-left in the panel in Cartesian coordinates.

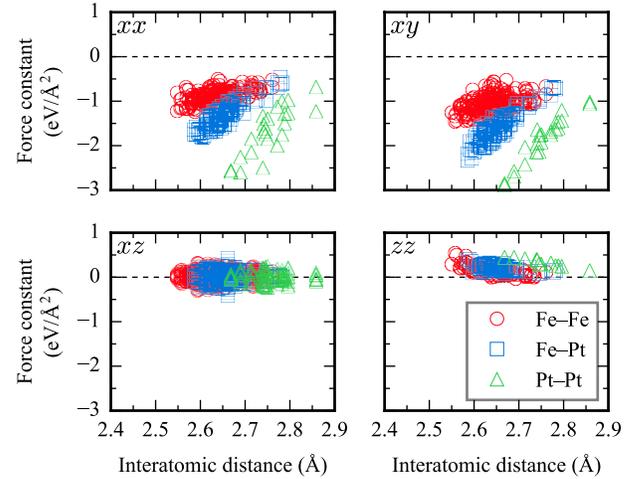

FIG. S3. The same as Fig. S2, but for disordered fcc $Fe_{0.75}Pt_{0.25}$ in the ideal FM state.

The FCs also depend on the interatomic distance even when the chemical pairs are the same. The FCs decrease in magnitude as the interatomic distance increases. Since the differences of the interatomic distance among the same chemical pairs should be caused by the differences of the local environments around the atomic pairs, the current result means that the FCs depend not only on their chemical pairs but also on their local environments. In this study the average element-resolved FCs are employed in the ICPA and the band-unfolding calculations, and therefore the local-environment dependence is not explicitly included for the calculated phonon spectra.

The FCs of Fe–Fe pairs are distributed more widely in the ideal PM state than in the ideal FM state. This is because in the PM state both the spin-up and the spin-down Fe are involved.

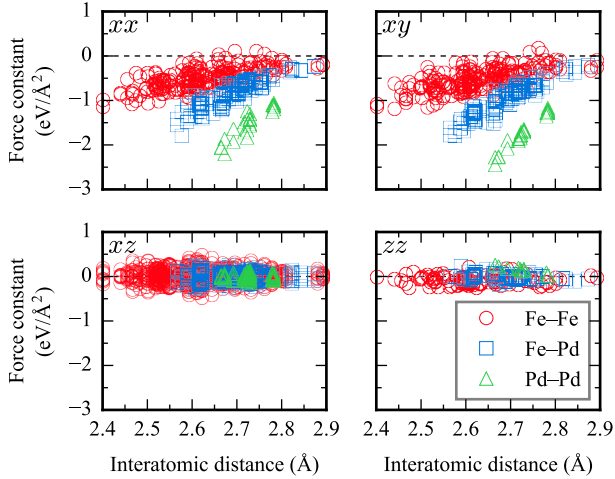

FIG. S4. The same as Fig. S2, but for disordered fcc $Fe_{0.75}Pd_{0.25}$ in the ideal PM state. Note that the magnetic moments on Fe is not distinguished in the plot.

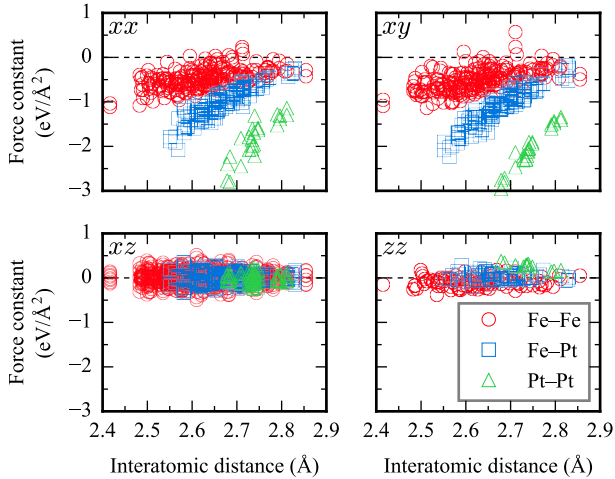

FIG. S5. The same as Fig. S4, but for disordered fcc $Fe_{0.75}Pt_{0.25}$ in the ideal PM state.

In contrast, the distributions of the FCs of Pd–Pd/Pt–Pt pairs for $Fe_{0.75}Pd_{0.25}$/$Fe_{0.75}Pt_{0.25}$ are similar between the ideal FM and the PM states.

### E. Supercell convergence of the unfolded phonon spectra

The phonon spectra calculated using the band unfolding depend on the size of the supercell model for which the phonon mode eigenvectors are calculated. Here we therefore investigate the supercell-size dependence of the unfolded phonon spectra. We considered three supercell models with different sizes, namely the 32-atom 2×2×2, the 256-atom 4×4×4, and the 846-atom 6×6×6 supercells of the conventional fcc unit

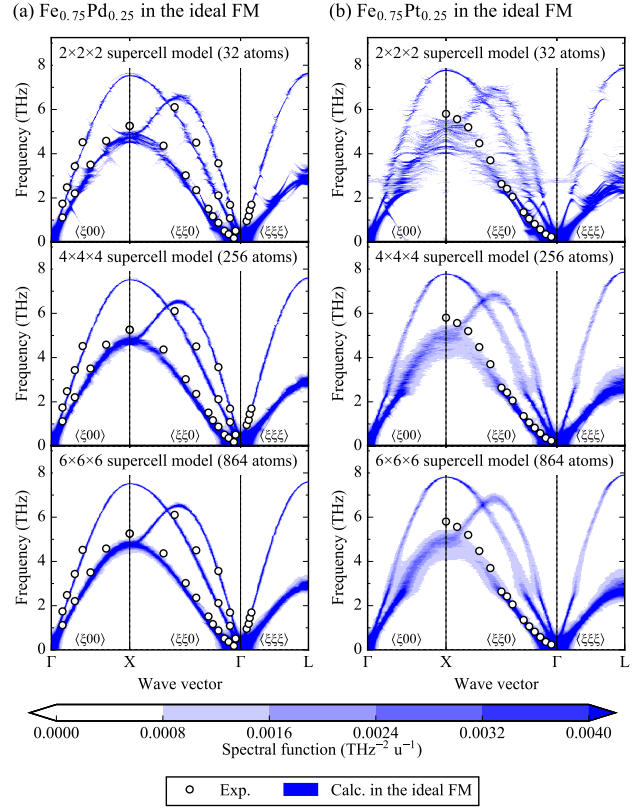

FIG. S6. Unfolded phonon spectra of $Fe_{0.75}Pd_{0.25}$ and $Fe_{0.75}Pt_{0.25}$ in the ideal FM state obtained from the supercell models with three different sizes. Color-contour corresponds to the relative magnitude of the spectral functions. White circles show the experimental phonon frequencies for disordered fcc $Fe_{0.72}Pd_{0.28}$ [S14] and $Fe_{0.72}Pt_{0.28}$ [S18] at room temperature, where these disordered alloys are in the FM phase.

cell. The atomic configurations of the 2×2×2 supercell models were the same as those of the SQS models used to extract the FCs, while the atomic configurations of the larger supercell models were determined using a pseudorandom number generator. Figure S6 shows the unfolded phonon spectra of the $Fe_{0.75}Pd_{0.25}$ and $Fe_{0.75}Pt_{0.25}$ in the ideal FM state obtained from these supercell models. Even for the smallest 2×2×2 supercell models (top panels), the phonon spectra are qualitatively similar to those obtained from the larger supercell models. The spectra of the 2×2×2 supercell models are, however, not smooth compared with those obtained from larger supercell models. The rugged spectra obtained from the 2×2×2 supercell models mainly originate from less number of sampled local environments around the atoms. In contrast, the spectra obtained from the 4×4×4 supercell models (middle panels) and from the 6×6×6 supercell models (bottom panels) are smooth and very similar to each other for both alloys. This implies that the spectra obtained from the 6×6×6 supercell models are probably almost converged with respect to the supercell size.

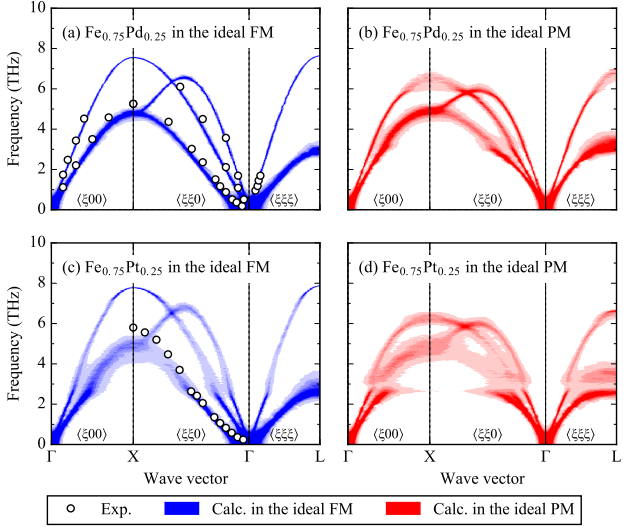

FIG. S7. Phonon spectra of $Fe_{0.75}Pd_{0.25}$ and $Fe_{0.75}Pt_{0.25}$ in the ideal FM and the ideal PM states calculated using the band unfolding. Color-contour corresponds to the relative magnitude of the spectral functions. White circles show the experimental phonon frequencies for disordered fcc $Fe_{0.72}Pd_{0.28}$ [S14] and $Fe_{0.72}Pt_{0.28}$ [S18] at room temperature, where these disordered alloys are in the FM phase.

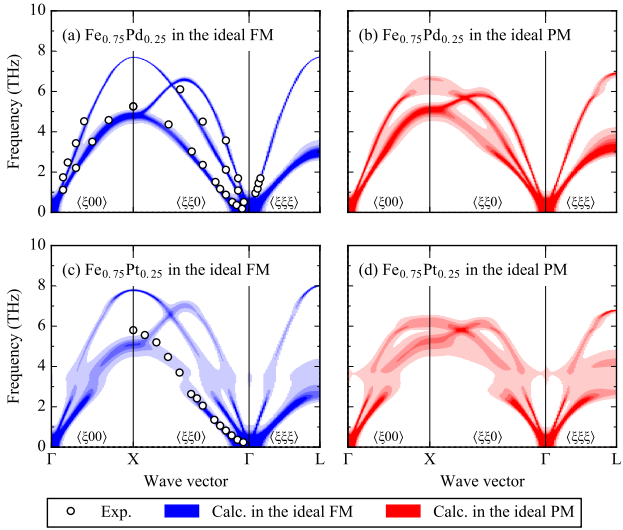

FIG. S8. The same as Fig. S7, but for the results calculated using the ICPA.

## F. Phonon spectra in the ideal FM and in the ideal PM states

Here the calculated phonon spectra of $Fe_{0.75}Pd_{0.25}$ and $Fe_{0.75}Pt_{0.25}$ in the ideal FM and the ideal PM states are compared. Figures S7 and S8 show the results obtained using the band unfolding [S16] and the itinerant coherent potential approximation (ICPA) [S17], respectively. Both methods are found to give very similar results for both alloys also in the ideal PM state.

The spectra in the ideal FM and the ideal PM states are largely different from each other for both alloys. When the magnetic state changes from the ideal FM to the ideal PM, the phonon frequencies of the lowest-frequency branch along the ⟨110⟩ direction tend to increase, while those of the relatively high-frequency branches tend to decrease. In addition, all the investigated phonon branches of $Fe_{0.75}Pt_{0.25}$ in the ideal PM state look discontinuous around 3 THz. We note that similar behaviors are found in the computed phonon dispersion relations for nonmagnetic $Cu_{0.75}Au_{0.25}$ [S16, S19] and $Ni_{0.50}Pt_{0.50}$ [S20].

## G. Temperature dependence of the phonon spectra calculated using the ICPA

Figure S9 shows the temperature dependence of the phonon spectra calculated using the ICPA for the same branch as that analyzed in Fig. 3 in the main text. The phonon frequencies calculated using the temperature-dependent FCs in Eq. (1) in the main text show good agreement with the experimental val-

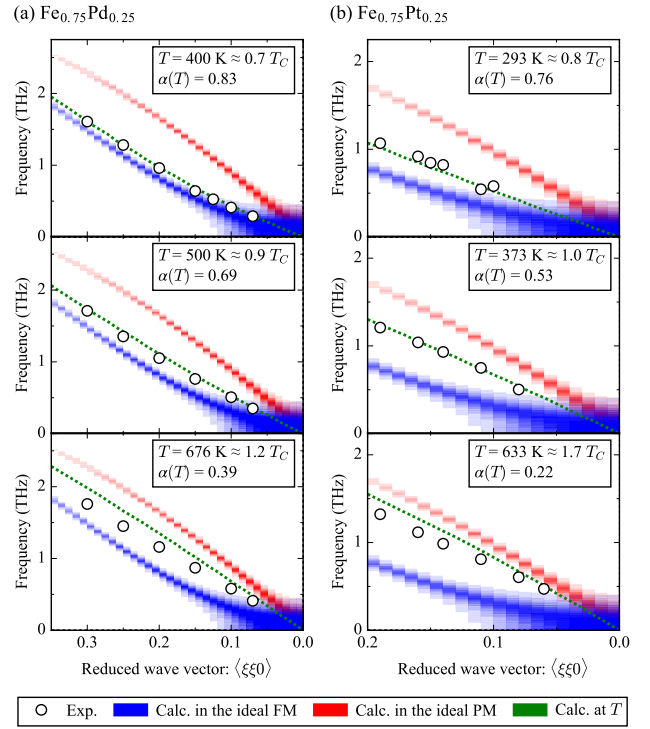

FIG. S9. Temperature dependence of the phonon spectra for the lowest-frequency branch along the ⟨110⟩ direction calculated using the ICPA. (a) $Fe_{0.75}Pd_{0.25}$. (b) $Fe_{0.75}Pt_{0.25}$. Green dotted curves show the peak positions of the phonon spectra obtained from the FCs for the corresponding temperatures in Eq. (1) in the main text. We also show the phonon spectra in the ideal FM and the ideal PM states as the blue and red color-contours, respectively, for comparison. $\alpha(T)$ used to obtain the FCs at the corresponding temperatures are also given in the panels. White circles show the experimental phonon frequencies for $Fe_{0.72}Pd_{0.28}$ [S14] and $Fe_{0.72}Pt_{0.28}$ [S21].

ues at the corresponding temperatures, as well as the phonon spectra obtained using the band unfolding.



\end{document}